\documentclass[fp]{jpsj3}
\usepackage{txfonts} 
\usepackage{bm}
\usepackage{amsmath}

\title{First principles study on the thermoelectric performance of CaAl$_2$Si$_2$-type Zintl phase compounds}

\author{Hidetomo Usui$^1$ and Kazuhiko Kuroki$^2$}
\inst{$^1$Department of Physics and Materials Science, 1060 Nishikawatsu-cho, Matsue, Shimane 690-8504, Japann\\
$^2$Department of Physics, 1-1 Machikaneyama-cho, Toyonaka, Osaka 560-0043, Japann} %\\

\abst{
We investigate the thermoelectric properties of CaAl$_2$Si$_2$-type Zintl phase compounds $AB_2X_2$ ($A$ = Mg, Ca, Sr, Ba, $B$ = Mg, Zn, Cd, and $X$ = P, As, Sb) using first principles band calculations within the Boltzmann transport theory assuming the constant relaxation time approximation.
We introduce the effective degree of valley degeneracy $n_{TE}$ to focus on the relationship between the thermoelectric properties and the multivalley character of the electronic band structure around the Fermi level.
We also introduce a quantity $\gamma_{TE}$, which takes into account $n_{TE}$ and anisotropy of the valley structure, and it is found that $\gamma_{TE}$  enables us to well understand the overall trend of the material dependence of the power factor.
We finally suggest promising thermoelectric materials, e.g. BaMg$_2$P$_2$ for PF $\sim 20\mu$W/cmK$^2$ and $ZT > 0.2$ at 300K and SrZn$_2$As$_2$ for PF $\sim 35\mu$W/cmK$^2$ and $ZT > 0.35$ at 300K assuming a relaxation time of 10 fs and a lattice thermal conductivity value of 2 W/mK.
}

\begin{document}
\maketitle

\section{Introduction}
Thermoelectric materials have recently attracted much attention owing to its capability of directly converting waste heat into electricity.
Conversion efficiency of thermoelectric converters is correlated with the dimensionless figure of merit $ZT$ given as follows,
\begin{eqnarray}
  ZT = \frac{\sigma S^2}{\kappa}T,
\end{eqnarray}
where $\sigma$ is the electrical conductivity, $S$ is the Seebeck coefficient, $\kappa$ is the thermal conductivity, and $T$ is the absolute temperature.
The power factor $PF=\sigma S^2$, which is the numerator of $Z$, is related to the power of the thermoelectric generation.
The thermal conductivity $\kappa$ is described as the summation of the lattice term $\kappa_{lat}$ and the electronic term $\kappa_{el}$.

In order to realize high thermoelectric efficiency, the thermoelectric materials should exhibit the coexistence of large power factor and small thermal conductivity.
Reducing the lattice thermal conductivity $\kappa_{lat}$ by nanostructuring is a popular way for obtaining large $ZT$ because nanostructuring can reduce the phonon relaxation time while keeping the carrier mobility\cite{Hochbaum2008, Boukai2008, Poudel2008}.
As for the power factor, it is generally difficult to increase both the electrical conductivity and the Seebeck coefficient at the same time due to the anti-correlation between the electrical conductivity and the Seebeck coefficient against the carrier concentration.

From the band structure point of view, a combination of high group velocity of electrons and large density of states around the Fermi level can result in large power factor.
Thus, specific band structures have been suggested such as multi-valley band\cite{Pei2011}, pudding-mold-type band\cite{Kuroki2007}, and low dimensional band structure\cite{Hicks1993_1, Hicks1993_2,Dresselhaus2007}.
Especially, the multi-valley structure can play an important role in realizing good thermoelectric performance through material designing such as elemental substitution because a relatively small change in the elements and/or the crystal structure can affect largely the valley degeneracy.\cite{Pei2011}
Recently discovered thermoelectric material $\alpha$-MgAgSb ($ZT > 1 $ at $\sim 550$ K)\cite{Liu2018} 
is another example of materials where multi-valley band structure enhances the thermoelectric efficiency.

Recently, a number of Zintl phase compounds have been discovered as good thermoelectric materials.\cite{Shuai2017}
For instance, Yb$_{14}$MnSb$_{11}$\cite{Brown2006}, Ca$_9$Zn$_{4.6}$Sb$_9$\cite{Ohno2017}, YbCd$_{1.6}$Zn$_{0.4}$Sb$_2$\cite{Wang2009} and Mg$_{3.2}$Sb$_{1.5}$Bi$_{0.49}$Te$_{0.01}$\cite{Tamaki2016} have been found to have large $ZT$ exceeding 1.
Yb(Cd,Zn)$_2$Sb$_2$ and Mg$_{3+\delta}$(Sb,Bi,Te)$_{2}$ are 1-2-2 type Zintl phase compounds $AB_2X_2$, where $A$ is alkaline earth metal or lanthanoid atom, $B$ is Mg, Zn, or Cd, and $X$ is pnictogen atom.
1-2-2 type Zintl phase compounds are crystallized in CaAl$_2$Si$_2$-, BaCu$_2$S$_2$- or ThCr$_2$Si$_2$-type structure.
For CaAl$_2$Si$_2$-type structure (the $P\bar{3}m1$ space group), both the n- and p-type thermoelectric compounds exhibiting $ZT > 1$ have been found.\cite{Wang2009, Guo2011, Tamaki2016, Zhang2010, Shuai2016, Shuai2016-2, JZhang2016, Shuai2017}
It has been pointed out that the thermoelectric properties of CaAl$_2$Si$_2$-type compounds can be understood within the multi-valley scenario.\cite{Zhang2016, Tamaki2016, Zhang2019}
For the p-type compounds, the valence band structure mainly consists of the $p$ orbitals of atoms at the $X$ site, and the number of the $p$ bands around the Fermi level strongly correlates with the power factor and the dimensionless figure of merit.
Also in the electron doped case, Mg$_3$Sb$_2$ possesses multi-valley structure, where the conduction band is minimized at around the L point in the Brillouin zone.

Both theoretically and experimentally, 1-2-2 type Zintl phase compounds for $X$=Sb have been widely investigated\cite{Wang2009, Guo2011, Tamaki2016, Zhang2010, Shuai2016, Shuai2016-2, JZhang2016, Shuai2017, Zhang2016,Sun2017,Peng2018, Toberer2010, Singh2013, Zhang2018, Pomrehn2014, Tani2010, Zhang2019}, but in the last few years, 1-2-2 type arsenides have been discovered as good thermoelectric materials.
For instance, K-doped BaZn$_2$As$_2$ and BaCd$_2$As$_2$ possess $ZT > 0.3$ and 0.8, respectively.\cite{Kihou2017, Kunioka2018, Kunioka2020}
The thermal conductivity for $X={\rm As}$ and ${\rm P}$ however tends to be larger than that for $X={\rm Sb}$, but small lattice thermal conductivity has been obtained for K-doped BaCd$_2$As$_2$ (0.46 W/mK at 773K) and K-doped BaZn$_2$As$_2$ ($\sim 1$W/mK at 773K).
The number of the valley structure of the phosphides and arsenides in the hole doped regime tend to be large due to the reduction of the spin-orbit coupling constant compared to the antimonides\cite{Zhang2016,Sun2017, Belfarh2018}, so that 
thermoelectric performance of the phosphides and arsenides is expected to become larger than that of antimonides.
Thus promising thermoelectric materials could be discovered in the 1-2-2 type Zintl phase arsenides and phosphides.

Given this background, we investigate the thermoelectric properties of 25 compounds of CaAl$_2$Si$_2$-type crystal structure in not only antimonides but also arsenides and phosphides.
In order to analyze the relationship between the multi-valley structure and the thermoelectric properties, we introduce a quantity $n_{TE}$ that stands for the effective number of valleys contributing to the thermoelectric properties, and also a quantity $\gamma_{TE}$ which corresponds to a combination of the magnitude of the anisotropy of the band structure and the number of valleys.
In the hole doped regime, the split of the $p$ bands due to spin-orbit coupling still appears not only in antimonides but also in arsenides and phosphides, and it is found that phosphides are suitable for maximizing the number of valleys.
The number of valleys in the electron doped regime tends to be larger than 3 because of the lowest conduction band at the M point.
The degree of the valley degeneracy is maximized at $A = {\rm Mg}$ or ${\rm Zn}$ and $B = {\rm Mg}$, in which the $s$- and $p$-orbital components at the $A$ and $B$ sites construct the conduction band structure.
We conclude that CaAl$_2$Si$_2$-type phosphides and arsenides can be promising thermoelectric materials, such as p-type BaMg$_2$P$_2$ for PF $\sim 20\mu$W/cmK$^2$ and $ZT > 0.2$ at 300K and n-type SrZn$_2$As$_2$ for PF $\sim 35\mu$W/cmK$^2$ and $ZT > 0.35$ at 300K assuming the relaxation time to be 10 fs and the lattice thermal conductivity value to be 2 W/mK.

\section{Method}
For the 25 compounds listed in Table \ref{table_sup1}, the crystal structure is optimized by means of first principles calculations with the VASP package\cite{VASP1,VASP2,VASP3,VASP4,VASP5} using the Perdew--Burke--Ernzerhof exchange-correlation functional revised for solids (PBEsol)\cite{PBEsol} without considering the spin-orbit coupling.
We take a $20 \times 20 \times 10$ $k$-mesh in the Brillouin zone, and plane wave cut-off energy of 816 eV.
Using the optimized crystal structures, the electronic band structure and the thermoelectric properties are performed using the WIEN2k package\cite{WIEN2k, WIEN2k2} with modified Becke--Johnson potential\cite{MBJ1,MBJ2} considering the spin-orbit coupling.
We set $RK_{max} = 8$, and use 2,000 and 100,000 $k$-points in the Brillouin zone for self-consistent calculation and calculation of the thermoelectric properties, respectively.

For calculating the thermoelectric properties, we use the BoltzTraP code\cite{BoltzTraP}, which can evaluate the Seebeck coefficient, the electrical conductivity, and the electronic term of the thermal conductivity within the Boltzmann transport theory assuming the constant relaxation time $\tau$. The thermoelectric properties using the BoltzTraP code are calculated as described below,
\begin{eqnarray}
  \bm{K}_m &=& \tau \sum_{n,\bm{k}} \bm{v}_{n}(\bm{k}) \otimes \bm{v}_{n}(\bm{k}) \left( -\frac{\partial f}{\partial \varepsilon_{n,\bm{k}}} \right) (\varepsilon_{n,\bm{k}} - \mu)^m, \\
  \bm{\sigma} &=& e^2 K_0, \\
  \bm{S} &=& -\frac{1}{eT}\bm{K}^{-1}_{0}\bm{K}_1, \\
  \bm{\kappa}_{\rm el} &=& \frac{1}{T}\left( \bm{K}_2 - \bm{K}_1\bm{K}_0^{-1}\bm{K}_1\right),
\end{eqnarray}
where $\bm{K}_m$ is the transport coefficient tensor, $e (> 0)$ is the elementary charge, $\bm{v}_{n}(\bm{k})$ is the group velocity of electrons at a certain wave vector $\bm{k}$ and the $n$-th band structure $\varepsilon_{n,\bm{k}}$, $f$ is the Fermi-Dirac distribution function, and $\mu$ is the chemical potential. Within the constant relaxation time approximation, the Seebeck coefficient does not depend on the relaxation time, but the electrical conductivity and the electronic thermal conductivity depend on the relaxation time.
Because it is difficult to evaluate the relaxation time and the thermal conductivity using first principles calculations, we assume $\tau=10$ fs and $\kappa_{lat}=2$ W/mK in this study.
%The value of the relaxation time can be suitable for the 1-2-2 type Zintl phase compounds, e.g. the relaxation time of BaZn$_2$As$_2$ is 12 fs at 300K.
In fact, it has been found that the lattice thermal conductivity is around 1 to 2 W/mK for 1-2-2 type arsenides BaCd$_2$As$_2$ and BaZn$_2$As$_2$.\cite{Kihou2017,Kunioka2018,Kunioka2020}

%We evaluate the effective mass using the wave vector in the energy range from 0.1 eV measured from the edge of the band at each point. This is because $O(k_BT)$ contribute to the thermoelectric properties. Therefore, the difference of the band structure within several $k_BT$ does not recognize for thermoelectric effect.

\section{Results and Discussion}
\subsection{The crystal structure and the electronic band structure}
\begin{figure}
  \begin{center} 
  \includegraphics[width=10cm]{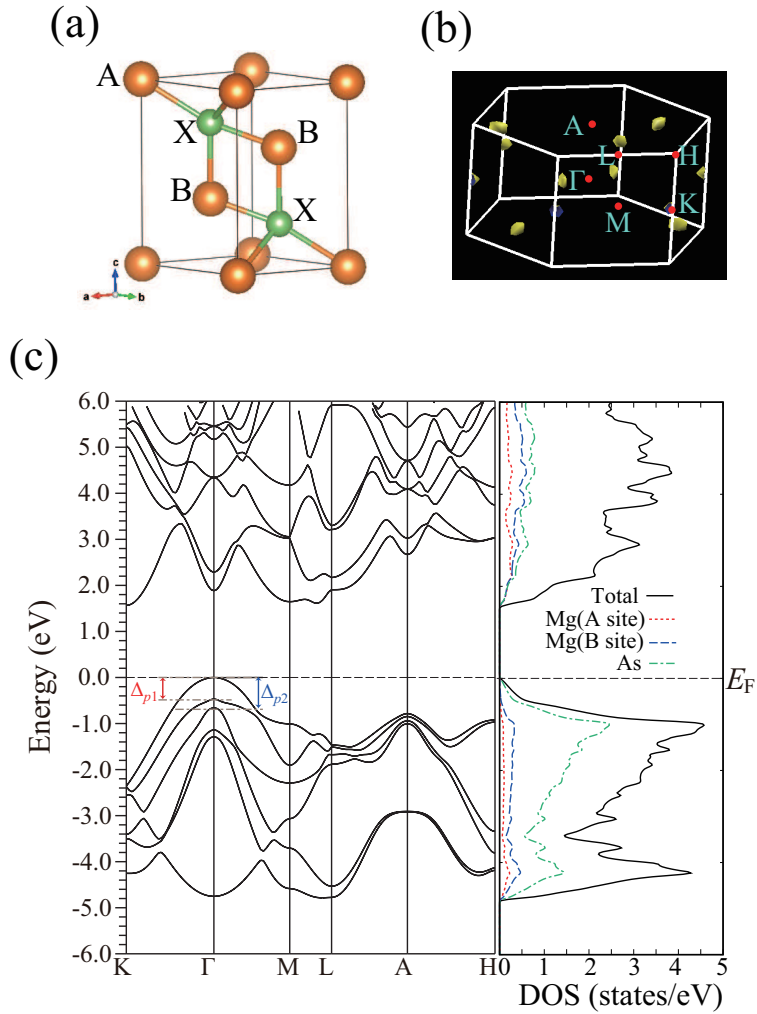}
  \caption{(a) The crystal structure depicted by VESTA\cite{VESTA}, (b) the Fermi surface and the Brillouin zone, and (c) the band structure and the density of states of Mg$_3$As$_2$. The Fermi surface is depicted at the energy of 0.07 eV measured from the bottom of the conduction band.}
  \label{fig1}
  \end{center}
\end{figure}

We first show the crystal structure of a CaAl$_2$Si$_2$-type compound Mg$_3$As$_2$ in Fig. \ref{fig1}(a).
Almost all the crystal structures and the band structures of $X={\rm Sb}$ have already been discussed in previous studies, so that we will mainly show the result of arsenides.
The CaAl$_2$Si$_2$-type crystal structure $AB_2X_2$ consists of alternately stacked [$B_2X_2$]$^{2-}$ anionic layers and $A^{2+}$ cation sheets.
In a [$B_2X_2$]$^{2-}$ two-dimensional layer, pnictogen atoms at the $X$ site form two edge sharing tetrahedra, where the atoms at the $B$ site are centered.
The optimized lattice parameters $a$ and $c$, and the internal coordinates at the $B$ and $X$ sites are listed in Table \ref{table1} for the arsenides and Table \ref{table_sup1} for the 25 compounds.
The calculated structural parameters are in good agreement with the experimental results.
For example, the crystal parameters of Mg$_3$As$_2$ are $a=4.26, c=6.73\mbox{\AA}$, $z_B=0.63$ and $z_X=0.23$ (experiment\cite{Mg3As2}), and $a=4.26$, $c=6.69\mbox{\AA}$, $z_B=0.642$ and $z_X=0.229$ (present calculation).

\begin{table}
  \begin{center} 
  \caption{The lattice parameters $a$ and $c$ (unit: $\mbox{\AA}$), and the internal coordinates $z_B$ for the $B$ site and $z_X$ for the $X$ site, obtained by the first principles calculations. The internal coordinates are $1a (0,0,0)$ for the $A$ site, $2d (1/3, 2/3, z)$ for the atoms of the $B$ and $X$ sites. \label{table1}}
\begin{tabular}{|c||c|c|c|c|c|c|c|} \hline
 compound & $a$ & $c$ & $z_B$ & $z_X$  \\ \hline \hline
  Mg$_3$As$_2$ & 4.26  & 6.69  & 0.642  & 0.229   \\ \hline
CaMg$_2$As$_2$ & 4.34  & 7.04  & 0.635  & 0.247   \\ \hline
CaZn$_2$As$_2$ & 4.13  & 6.92  & 0.631  & 0.257   \\ \hline
CaCd$_2$As$_2$ & 4.38  & 7.09  & 0.633  & 0.234   \\ \hline
ZnMg$_2$As$_2$ & 4.23  & 6.49  & 0.646  & 0.218   \\ \hline
SrMg$_2$As$_2$ & 4.40  & 7.34  & 0.630  & 0.260   \\ \hline
SrZn$_2$As$_2$ & 4.19  & 7.19  & 0.629  & 0.270   \\ \hline
SrCd$_2$As$_2$ & 4.44  & 7.36  & 0.632  & 0.249   \\ \hline
BaMg$_2$As$_2$ & 4.47  & 7.68  & 0.625  & 0.272   \\ \hline
BaCd$_2$As$_2$ & 4.51  & 7.64  & 0.630  & 0.262   \\ \hline
 \end{tabular}
\end{center}
\end{table}

The electronic band structure of the 25 compounds is shown in Figs. \ref{fig_supp1}, \ref{fig_supp2} and \ref{fig_supp3}. 
Due to the $P\bar{3}m1$ space group, the Brillouin zone shown in Fig. \ref{fig1}(b) has a hexagonal prism structure, and there are one $\Gamma$ point, two K/H points, and three M/L points.
The electronic band structure and the density of states for Mg$_3$As$_2$ shown in Fig. \ref{fig1}(c) clearly indicate that the six valence bands ranging from $-6$ to 0 eV measured from the Fermi level mainly consist of the $p$ orbitals of the As atoms at the $X$ sites.
The highest valence band mainly originates from the $p_z$ orbital, and the second and third highest bands are mainly composed of the hybridization of the $p_x$ and $p_y$ orbitals.
Due to the symmetry of the crystal structure and the orbital character of the $p_x$ and $p_y$ orbitals, the second and third highest bands should have two-fold degeneracy at the $\Gamma$ point, but the hybridization between all the $p$ orbitals due to spin-orbit coupling split the two bands.
In the conduction bands, the band structure originates from the hybridization of orbitals of all the atoms, but is mainly composed of the $s$ orbital of the Mg atoms positioned at the $A$ site and the $s$ and $p$ orbitals at the $B$ site.
The bottom of the conduction band is positioned at the K point and a certain wave vector around the L point, which corresponds to eight Fermi surfaces in the electron doped regime as shown in Fig. \ref{fig1}(b).
The eight valleys in the conduction band thus contribute to the thermoelectric effect in the electron doped case.

\subsection{Thermoelectric properties}

\begin{figure}
  \begin{center} 
  \includegraphics[width=10cm]{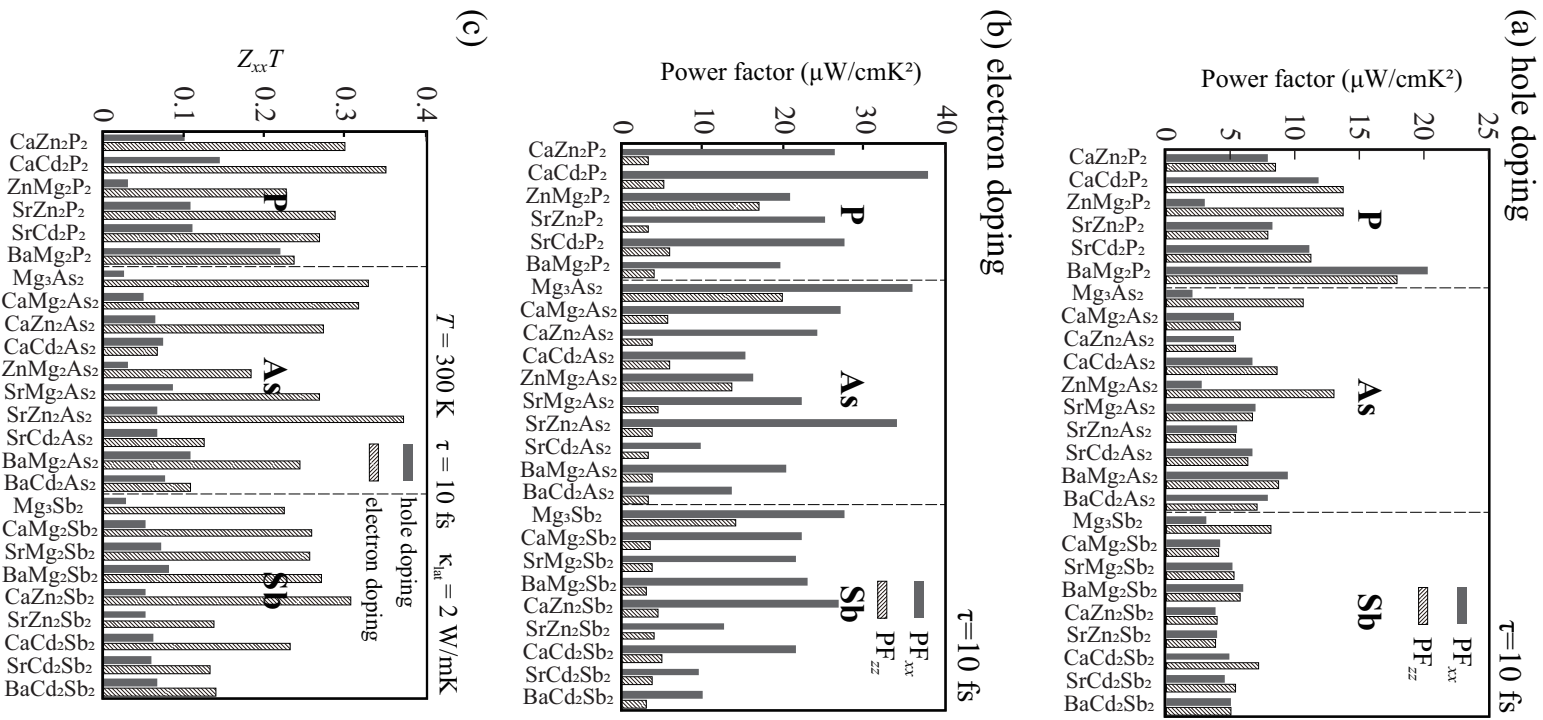}
  \caption{The maximum values of (a),(b) the power factor and (c) the dimensionless figure of merit of each calculated compound at 300K, $\tau = 10$fs, and $\kappa_{lat}=2{\rm W/mK}$.}
  \label{fig5}
  \end{center}
\end{figure}

In Figs. \ref{fig5}(a) and (b), we show the maximum $xx$ and $zz$ components of the power factor, within the doping level of 0.1 carriers per unit cell, for each compound in the hole and electron doped regime. 
The doping level at which the maximum power factor is obtained is listed in Tables \ref{table_sup2} and \ref{table_sup3} within this doping level.
Actually, some compounds e.g. Mg$_3$As$_2$ exhibit large power factor even at around 0.5 holes per unit cell because of the large density of states at around $-1$ eV (Fig. \ref{fig1}(c)), but the Seebeck coefficient is smaller than 50 $\mu$V/K, which does not result in a large figure of merit exceeding 1.
This is because the maximum value of the dimensionless figure of merit is determined by the ratio of $S^2$ to the Lorenz number $L$ when we assume the Wiedemann--Franz law to hold and $\kappa_{lat}=0$, and at least $S \sim 160$ $\mu$V/K is required to obtain $ZT \sim 1$.

In the hole doped regime, the maximum value of the $xx$ and $zz$ components of the power factor is comparable to each other (Fig. \ref{fig5}(a)) because the effective mass in the $x$, $y$ and $z$ directions is comparable, e.g. for CaZn$_2$As$_2$, the effective mass of the first valence band is $m_{x}=-0.42m_e, m_{y}=-0.58m_e, m_{z}=-0.61m_e$, where $m_e$ is the electron mass.
The maximum power factor for the P system tends to be larger than that for the As and Sb systems, whereas the power factor in the electron doped regime (Fig.\ref{fig5}(b)) does not strongly depend on the pnictogen atom at the $X$ site.
The maximum values of the dimensionless figure of merit along the $x$-axis is shown in Fig. \ref{fig5}(c).
The result of $ZT$ reflects the tendency of the power factor in the electron and hole doped regime, namely, p-type $ZT$ in the P system tends to be larger than that in the As and Sb systems, assuming $\kappa_{lat} = 2 $W/mK for all the compounds.

\subsection{Valley degeneracy in the hole doped regime}

In the hole doped regime, the valley degeneracy originating from the three bands around the Fermi level plays an important role for the thermoelectric effect.
In Ref. \citen{Zhang2016}, the energy difference $\Delta_p$ has been used to reflect the degeneracy of the three bands around the Fermi level.
$\Delta_p$  is defined as  $\Delta_p = E_{p_{x/y}} - E_{p_z}$, where $E_{p_{x/y}}$ and $E_{p_z}$ are the energy of the two $p_{x/y}$ bands and one $p_z$ band at the $\Gamma$ point.
In this evaluation, the two $p_{x/y}$ bands are degenerate at the $\Gamma$ point because the spin-orbit coupling is not considered.
As shown in Fig. \ref{fig1}(c), the spin-orbit coupling forces the $p_{x/y}$ bands to split, so that two energy differences $\Delta_{p1}$ and $\Delta_{p2}$ should be considered for evaluating the valley degeneracy. $\Delta_{p1}$ and $\Delta_{p2}$ are the energy difference between the first and second highest valence bands, and between the first and third bands at the $\Gamma$ point, respectively (see Fig. \ref{fig1}(c)).
It has been found that $\Delta_{p1}$ is almost the same as $\Delta_{p}$, and $\Delta_{p2}$ is much larger than $k_BT$ in the Sb system\cite{Zhang2016}.
$\Delta_p$ can therefore reflect the degree of degeneracy of the first and second bands.

However, when both $\Delta_{p1}$ and $\Delta_{p2}$ are smaller than several $k_BT$, it is important to understand how many bands contribute to the thermoelectric properties using $\Delta_{p1}$ and $\Delta_{p2}$.
We thus introduce $n_{TE}^{hole}$,
the effective degree of the band degeneracy, extended to non-integer numbers, as described below,
\begin{eqnarray}
  n^{\rm hole}_{TE} = 4k_BT \sum^{2}_{i=0} \left( -\frac{df}{dE} \right)_{E=\Delta_{pi}/2}, \label{eq:n_hole}
\end{eqnarray}
where $\Delta_{p0}$ is zero, namely, the energy of the highest valence band at the $\Gamma$ point, and 
the chemical potential $\mu$ and the temperature $T$ that enter $df/dE$ are fixed at $\mu=0$ and $T=300$K
In Eq. (\ref{eq:n_hole}), $4k_BT$ is the normalization factor for obtaining $n^{\rm hole}_{TE} = 1 $ at $\Delta_{pi} = 0$.

Eq. (\ref{eq:n_hole}) originates from the transport coefficient $K_n$, which are also written as follows,
\begin{eqnarray}
\bm{K}_m &=& \int \bm{\sigma}(E)\left(-\frac{df}{dE}\right)(E-\mu)^m, \label{eq:K_n_int}\\
\bm{\sigma}(E) &=& \tau \sum_{n,\bm{k}} \bm{v}_{n}(\bm{k}) \otimes \bm{v}_{n}(\bm{k}) \delta(E-\varepsilon_{n,\bm{k}}). \label{eq:sigma}
\end{eqnarray}
$(-df/dE)$ in the transport coefficient $K_m$ extracts the energy range in which the band structure contribute to the thermoelectric effect. 
The contribution to the electrical conductivity thus comes from the band structure within the energy range between $\pm$ several $k_BT$.
For the Seebeck coefficient, due to the term of $(E-\mu)$ of $\bm{K}_1$ in Eq. (\ref{eq:K_n_int}), the energy range is wider than that for the electrical conductivity.
We use the energy at the $\Gamma$ point and $(-df/dE)_{E=\Delta_{pi}/2}$ for simplification instead of $(-df/dE)(E-\mu)^m$ to evaluate the contribution of the band structure to the thermoelectric effect.
The reason why we take $\mu=0$ in $df/dE$ is because we intend to analyze the correlation between $n_{\rm TE}^{\rm hole}$ and the maximum power factor (within the hole doping rate of $<0.1$), and the power factor tends to be maximized at a doping rate corresponding to $\mu \simeq 0$. In some cases, the power factor is maximized at $\mu$ away from zero, but even in this case, we have checked that the power factor at $\mu \simeq 0$ is comparable to its maximum value.
Using Eq. (\ref{eq:n_hole}), we obtain the effective degree of the valley degeneracy of 0.8, 0.4 and 0.1 for $\Delta_{pi}=2k_BT$, $4k_BT$, and $8k_BT$,	respectively. 

\begin{table}
  \begin{center} 
  \caption{The energy difference $\Delta_{p1}$ between the first and second highest band, and $\Delta_{p2}$ between the first and third highest bands (unit: eV), the effective degree of the valley degeneracy contributing to the thermoelectric effect $n^{\rm hole}_{TE}$ at 300K, The maximum value of the $xx$ component of the power factor $PF_{xx}$ (unit: $\mu$W/cmK$^2$) obtained with the relaxation time to be 10 fs within the doping level of less than 0.1 holes at 300K, the energy difference $\Delta_p$ between the $p_{x/y}$ and $p_{z}$ bands without considering the spin-orbit coupling, and the effective degree of the band degeneracy $n_{TE}^{hole0}$ using $\Delta_{p}$. \label{table3}}
  \begin{tabular}{|c|c|c|c|c|c|c|} \hline
          & \multicolumn{4}{|c|}{SOC}  & \multicolumn{2}{|c|}{no SOC}  \\ \cline{2-7}
          & $\Delta_{p1}$ & $\Delta_{p2}$ & $n^{\rm hole}_{TE}$ & $PF_{xx}$ &$\Delta_p$ & $n^{\rm hole0}_{TE}$ \\ \hline \hline
          Mg$_3$As$_2$   & 0.47 & 0.66 & 1.00 & 2.0  & -0.52 & 1.00 \\ \hline
          CaMg$_2$As$_2$ & 0.15 & 0.37 & 1.20 & 5.3  & -0.19 & 1.19 \\ \hline
          CaZn$_2$As$_2$ & 0.06 & 0.31 & 1.74 & 5.3  & 0.10 & 2.43 \\ \hline
          CaCd$_2$As$_2$ & 0.06 & 0.30 & 1.71 & 6.7  & 0.10 & 2.46 \\ \hline
          ZnMg$_2$As$_2$ & 0.74 & 0.93 & 1.00 & 2.8  & -0.81 & 1.00 \\ \hline
          SrMg$_2$As$_2$ & 0.03 & 0.27 & 1.95 & 6.9  & -0.04 & 2.75 \\ \hline
          SrZn$_2$As$_2$ & 0.08 & 0.34 & 1.57 & 5.5  & 0.15 & 2.18 \\ \hline
          SrCd$_2$As$_2$ & 0.07 & 0.30 & 1.68 & 6.7  & 0.11 & 2.38 \\ \hline
          BaMg$_2$As$_2$ & 0.00 & 0.26 & 2.02 & 9.5  & 0.01 & 2.99 \\ \hline
          BaCd$_2$As$_2$ & 0.07 & 0.31 & 1.68 & 7.9  & 0.11 & 2.37 \\ \hline
          BaMg$_2$Sb$_2$ & 0.07 & 0.61 & 2.00 & 4.2 & 0.01 & 2.99 \\ \hline
          BaMg$_2$P$_2$  & 0.01 & 0.08 & 2.59 & 20.3 & 0.00 & 3.00 \\ \hline
   \end{tabular}
  \end{center}
 \end{table}

\begin{figure}
  \begin{center}
  \includegraphics[width=10cm]{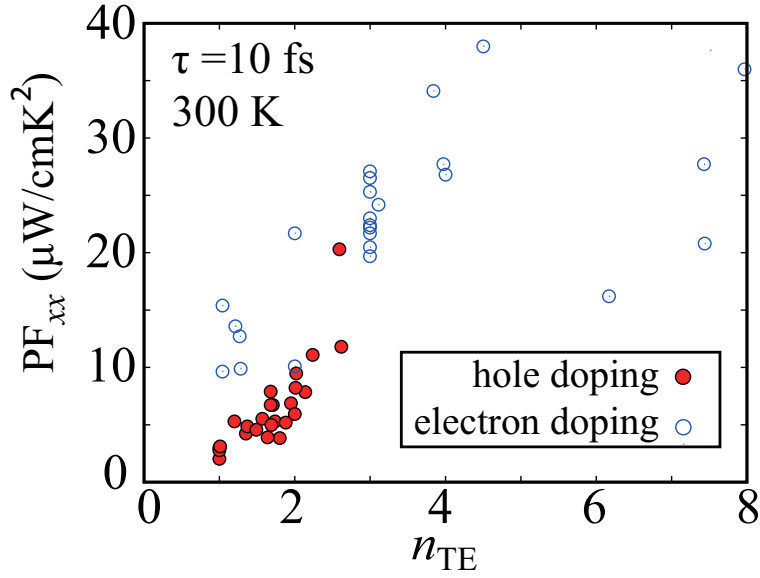}
  \caption{The maximum value of the $xx$ component of the power factor plotted against the effective degree of the band degeneracy contributing to the thermoelectric effect $n_{TE}$.}
  \label{fig3_1}
  \end{center}
\end{figure}

Table \ref{table3} shows, for the As systems (and also for BaMg$_2$Sb$_2$ and BaMg$_2$P$_2$),  the energy differences $\Delta_{p1/2}$, the maximum power factor within the doping level of less than 0.1 holes, and the effective degree of the valley degeneracy $n_{TE}^{hole}$.
The data for all the 25 compounds are listed in Table \ref{table_sup1}. 
The maximum value of the $xx$ component of the power factor is plotted against $n^{hole}_{TE}$ in Fig. \ref{fig3_1}. 
We can clearly see the correlation between the degree of the band degeneracy and the power factor in the hole doped regime.

\begin{figure}
  \begin{center}
  \includegraphics[width=10cm]{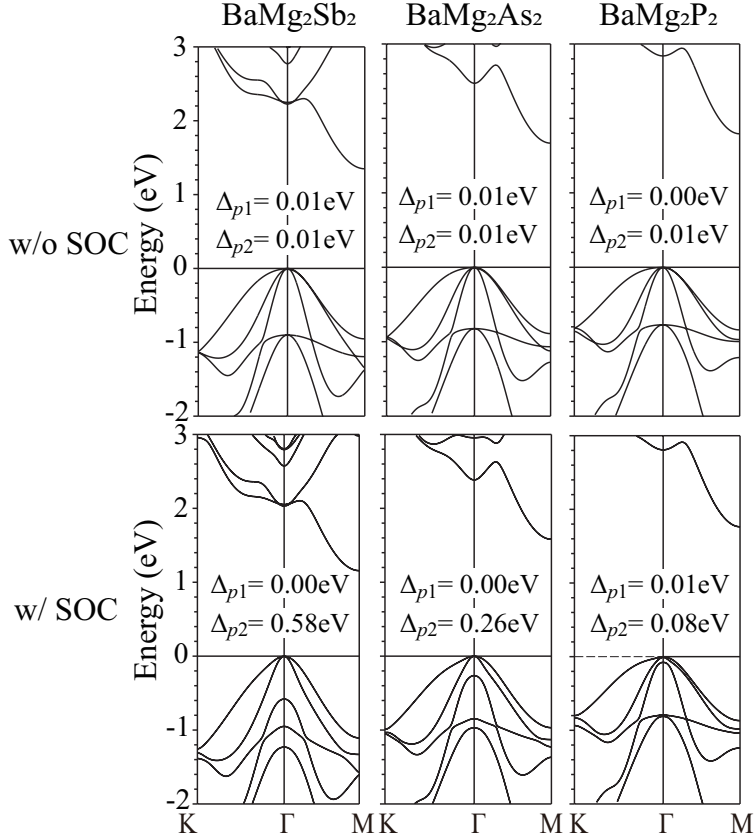}
  \caption{The electronic band structure of BaMg$_2$Sb$_2$, BaMg$_2$As$_2$ and BaMg$_2$P$_2$ calculated with and without considering spin-orbit coupling considered.}
  \label{fig2}
  \end{center}
\end{figure}

To understand the material dependence of the degree of degeneracy, we show in Fig. \ref{fig2} the electronic band structure calculated with or without the spin-orbit coupling considered.
When we ignore the spin-orbit coupling, the three $p$ bands of BaMg$_2X_2$ are almost degenerate at the $\Gamma$ point, but $\Delta_{p2}$ drastically increases when the spin-orbit coupling is turned on for the Sb and As systems.
The smallest value of $\Delta_{p2}$ for each pnictogen system is 0.58 for $X={\rm Sb}$, 0.26 for $X={\rm As}$, and 0.07 eV for $X={\rm P}$, which corresponds to 22$k_BT$ for Sb, 10$k_BT$ for As and 2.7$k_BT$ for P at 300K.
Thus, the effect of the spin-orbit coupling on $\Delta_{pi}$ is not so small even in the P system.
We evaluate the ratio of $n^{\rm hole}_{TE}$ with and without including spin-orbit coupling, and the median of the ratio is 0.68, 0.72 and 0.95 in the Sb, As and P systems, respectively.
Hence the P system is favorable at least from the viewpoint of enhancing the thermoelectric efficiency through increasing the degree of valley degeneracy. 

\begin{figure}
  \begin{center}
  \includegraphics[width=10cm]{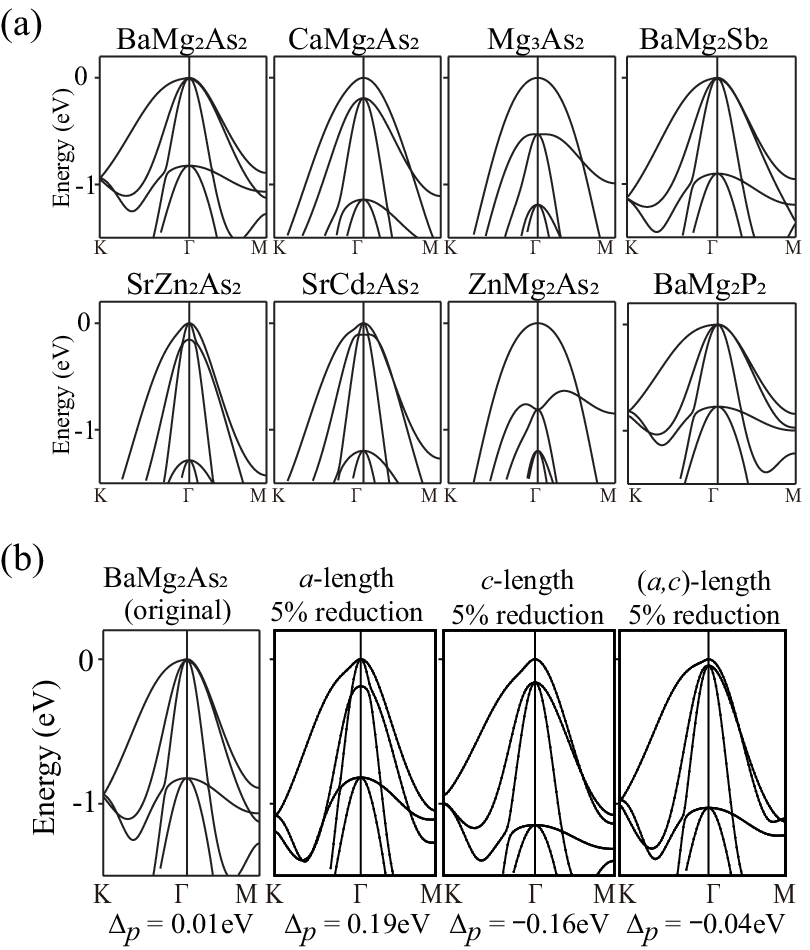}
  \caption{(a) The electronic band structure of $AB_2X_2$ without considering the spin-orbit coupling and (b) the electronic band structure of BaMg$_2$As$_2$ with original and modified lattice constants.}
  \label{fig3}
  \end{center}
\end{figure}

We move onto the discussion on the element dependence of the valley degeneracy.
Fig. \ref{fig3}(a) shows the valence band structure calculated without including the spin-orbit coupling.
It is found that the energy splitting between the $p_{x/y}$ and $p_{z}$ orbitals becomes larger with decreasing the atomic number at the $A$ site from Ba to Ca and Mg (see the band structure of $A$Mg$_2$As$_2$ for $A = {\rm Ba, Ca}$ and ${\rm Mg}$ in Fig. \ref{fig3}(a)).
We can naturally understand the $A$-site dependence of the energy splitting by considering the variation of the lattice parameters. 
Fig. \ref{fig3}(b) shows the band structure of BaMg$_2$As$_2$ with hypothetical crystal structures where the lattice parameters are reduced from the optimized ones and the internal coordinates are fixed at the optimized ones.
It is found that the $a$-length controls the As-As distance at the same $z$-coordinate in the [Mg$_2$As$_2$]$^{2-}$ layers, and thus the on-site energy and the hopping integrals of the $p_{x}$ and $p_{y}$ orbitals in the two-dimensional layers increase. 
The energy of the $p_{x/y}$ bands at the $\Gamma$ point becomes larger than that of the $p_z$ band when the $a$-length is reduced.
On the other hand, when the lattice parameter $c$ decreases, the energy of the $p_z$ band at the $\Gamma$ point increases.
Actually, substituting the Mg atom for the Ba atom at the $A$ site reduces the lattice constant $a$ and $c$ by 5\% and 13\%, respectively.
The correlation between the $(a,c)$-length and $\Delta_p$ can be seen in ZnMg$_2$As$_2$, where the lattice constant $c$ is smaller than that of Mg$_3$As$_2$.

This correlation can be seen in the elemental substitution at another site.
In the comparison of the lattice parameters of SrZn$_2$As$_2$ and SrCd$_2$As$_2$, the ratios of $a$ to $c$ are 0.58 and 0.60, respectively.
Thus the energy difference of the $p_{x/y}$ and $p_{z}$ bands for SrCd$_2$As$_2$ is a bit smaller than that for SrZn$_2$As$_2$. 
In the case of the elemental substitution at the $X$ site, the ratio of $a$ to $c$ barely changes from 0.58 to 0.59, so that the energy difference $\Delta_p$ remains almost unchanged against the variation of the pnictogen atom.
In fact, the energy difference of BaMg$_2X_2$ for $X = {\rm Sb, As, P}$ is nearly 0 eV when we exclude the spin-orbit coupling as shown in Fig. \ref{fig2}.
We can conclude that two of the important factors for increasing the effective degree of the valley degeneracy are (i) to increase the atomic number of the $A$ site for optimizing the lattice parameters, (ii) to reduce the atomic number of the $X$ site for reducing the spin-orbit coupling constant.
It is noted that the material dependence of $\Delta_p$ cannot be understood from the variation of $a/c$ alone.
For instance, $a/c$ of BaMg$_2$As$_2$ and SrZn$_2$As$_2$ is comparable but the difference in $\Delta_p$ between the two compounds is 0.15 eV.
This may be because of the difference in the group of elements at the $B$ site, which can affect the $B$-$X$ bonding condition.

We note that one of the $p_{x/y}$ bands along the $\Gamma$-M direction becomes less dispersive when we replace Ba with Mg at $A$ site (see the second highest valence band of Mg$_3$As$_2$ and BaMg$_2$As$_2$ in Fig. \ref{fig3}(a)).
Thus the effective mass of this band in Mg$_3$As$_2$ is much larger than that of BaMg$_2$As$_2$.
For ZnMg$_2$As$_2$, the effective mass of the $p_{x/y}$ band at the $\Gamma$ point is positive along the $\Gamma$-M direction, so that the multiple Fermi surfaces appear when $\mu$ is positioned at around $-0.7$ eV although this would be a heavily hole doped situation.
ZnMg$_2$As$_2$ could possess largest $n^{hole}_{TE}$ if the $p_z$ band sank below the $p_{x/y}$ bands.
From the viewpoint of the anisotropy of the electronic band structure, ZnMg$_2$As$_2$ and Mg$_3$As$_2$ also possess strongly anisotropic $p_{x/y}$ band, which results in the large density of states and the large group velocity. 
If this anisotropic feature of the band structure can be realized in the large $c$-length regime, where the top of the $p_{x/y}$ bands is above that of the $p_{z}$ band, large thermoelectric performance is expected.

\subsection{Valley degeneracy in the electron doped regime}

\begin{table}
  \begin{center} 
  \caption{The effective degree of the valley degeneracy contributing to the thermoelectric effect $n^{\rm ele}_{TE}$ in the electron doped regime at $\Gamma$, K, M point at $T= 300$K, The maximum value of the $xx$ component of the power factor $PF_{xx}$ (unit: $\mu$W/cmK$^2$) obtained with the relaxation time to be 10 fs within the doping level of less than 0.1 electrons at 300K. $\Gamma$1 and $\Gamma$2 show the first and second bands at the $\Gamma$ point, respectively. M$^*$  shows the bottom of the conduction band positioned at around the L point. \label{table2}}
\begin{tabular}[t]{|c|c|c|c|c|c|c|} \hline
 compound  & $\Gamma$1 & $\Gamma$2 &  K & M & $n^{\rm ele}_{TE}$ & $PF_{xx}$\\ \hline \hline
   Mg$_3$As$_2$ & 0.01  & 0.00  &  2.00  & 5.96 (M$^*$) & 7.97  & 36.0 \\ \hline
 CaMg$_2$As$_2$ & 0.00  & 0.00  &  0.00  & 3.00  & 3.00  & 27.1 \\ \hline
 CaZn$_2$As$_2$ & 0.11  & 0.00  &  0.00  & 3.00  & 3.11  & 24.2 \\ \hline
 CaCd$_2$As$_2$ & 1.00  & 0.03  &  0.00  & 0.01  & 1.04  & 15.4 \\ \hline
 ZnMg$_2$As$_2$ & 0.00  & 0.00  &  0.17  & 6.00 (M$^*$) & 6.17  & 16.2 \\ \hline
 SrMg$_2$As$_2$ & 0.00  & 0.00  &  0.00  & 3.00  & 3.00  & 22.4 \\ \hline
 SrZn$_2$As$_2$ & 0.84  & 0.00  &  0.00  & 3.00  & 3.84  & 34.1 \\ \hline
 SrCd$_2$As$_2$ & 1.00  & 0.28  &  0.00  & 0.00  & 1.28  & 9.9 \\ \hline
 BaMg$_2$As$_2$ & 0.00  & 0.00  &  0.00  & 3.00  & 3.00  & 20.5 \\ \hline
 BaCd$_2$As$_2$ & 1.00  & 0.20  &  0.00  & 0.01  & 1.21  & 13.6 \\ \hline
 \end{tabular}
 \end{center}
\end{table}

\begin{figure}
  \begin{center}
  \includegraphics[width=10cm]{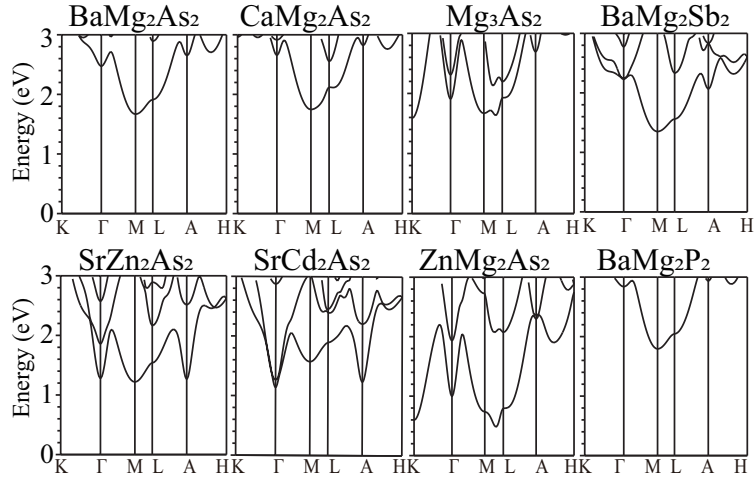}
  \caption{The conduction band structure of CaAl$_2$Si$_2$-type Zintl phase compounds obtained without considering the spin-orbit coupling.}
  \label{fig4}
  \end{center}
\end{figure}

We now discuss the effective degree of degeneracy in the electron doped regime.
Although the spin-orbit coupling reduces the degree of the valley degeneracy in the hole doped regime as shown in Fig. \ref{fig2}, the spin-orbit coupling barely affect the conduction bands near the Fermi level.
This is because the energy splitting due to the spin-orbit coupling does not exist in the lowest conduction bands.
Actually, the maximum power factor obtained considering the spin-orbit coupling shown in Fig. \ref{fig5} is almost the same as that obtained without considering the spin-orbit coupling (not shown).

We show the electronic band structure obtained without considering the spin-orbit coupling in Fig. \ref{fig4}.
The lowest conduction band is minimized at $\Gamma$, K or M point in the calculated compounds except for a few compounds such as Mg$_3X_2$.
To discuss the band degeneracy in the electron doped regime, we modify the expression of Eq. (\ref{eq:n_hole}) by including the band structure at other $k$-points, which is described below,
\begin{eqnarray}
  n^{\rm ele}_{TE} = 4k_BT  \sum_{i, \bm{k}} n_{\bm{k}} \left( -\frac{df}{dE} \right)_{E=\Delta_{pi\bm{k}}/2}, \label{eq:n_ele}
\end{eqnarray}
where $n_{\bm{k}}$ is the number of equivalent $k$-points for the wave vectors $\bm{k} = \Gamma$, M, and K ($n_{\bm{k}} = 1$ for $\bm{k}=\Gamma$, 2 for K, and 3 for M), and $\Delta_{pi\bm{k}}$ is the energy difference between the first and $i$-th lowest conduction band at the wave vector $\bm{k}$. $\mu$ and $T$ for $df/dE$ in Eq. (\ref{eq:n_ele}) are zero (bottom of the conduction band) and 300 K, respectively.
In Mg$_3$X$_2$ and ZnMg$_2$P$_2$, the conduction band minimum is positioned at around L point, so that we add this $k$-point denoted as ``M$^*$" in Table \ref{table2} in the summation of Eq. (\ref{eq:n_ele}) instead of the $M$ point.
The calculated degree of the valley degeneracy is shown in Table \ref{table2}, and the power factor is plotted against $n_{TE}$ in Fig. \ref{fig3_1}.
Also in the electron doped regime, positive correlation between the valley degeneracy and the power factor except for Mg$_3X_2$ and ZnMg$_2X_2$ exhibiting large $n_{TE}$ (see Tables \ref{table2}, and \ref{table_sup1}--\ref{table_sup3}).

In Table \ref{table2}, it is found that the degree of the valley degeneracy tends to be larger than three because the lowest energy of the conduction band of almost all the 25 compounds is minimized at the M point. The average value of $n^{\rm ele}_{TE}$ is actually 3.55, which means that the n-type CaAl$_2$Si$_2$-type compounds can exhibit larger thermoelectric performance than p-type compounds.
In addition, for $A = {\rm Mg}$ or ${\rm Zn}$, the conduction band minimum is positioned at around L point, so that the band degeneracy drastically increases due to the six-valley band structure. 

The six valleys of the lowest conduction band at the M$^*$ point may originate from the orbital character of the element at the $A$ site.
The conduction band minimum at the M$^*$ point appears only when the $p$ orbital at the $A$ site construct the conduction bands.
$A = {\rm Mg}$ or ${\rm Zn}$ fullfil this condition, and actually exhibit the six-valley structure in the conduction bands.
For $A = {\rm Ba, Sr}$, or ${\rm Ca}$, the $d$ orbitals of the $A$ atom construct the conduction bands, which are minimized at the M or $\Gamma$ point.
We calculated the electronic band structure of BaMg$_2$As$_2$ using the lattice parameters and the internal coordinates modified to be the same as those of Mg$_3$As$_2$, but the six-valley structure does not appear, so that 
we conclude that the six-valley structure does not originate from a peculiar set of lattice parameters and internal coordinates.
From the viewpoint of the band degeneracy, ZnMg$_2X_2$ and Mg$_3$As$_2$ can be promising thermoelectric materials.
For SrCd$_2$As$_2$, the lowest energy at the $\Gamma$ point is larger than that at the M point, so that 
the degree of the valley degeneracy therefore changes from 3 to 1 for the elemental substitution of $B$ = Zn with Cd.
The partial elemental substitution of Zn with Cd can thus maximize the band degeneracy at the Fermi level.
Table \ref{table2} shows that the number of valleys does not strongly depend on the pnictogen atom, but the average of the $xx$ component of the power factor increases when Sb is replaced with P and As, because of the increase of the effective mass.

\subsection{Valley degeneracy and anisotropy}

\begin{figure}
  \begin{center}
  \includegraphics[width=10cm]{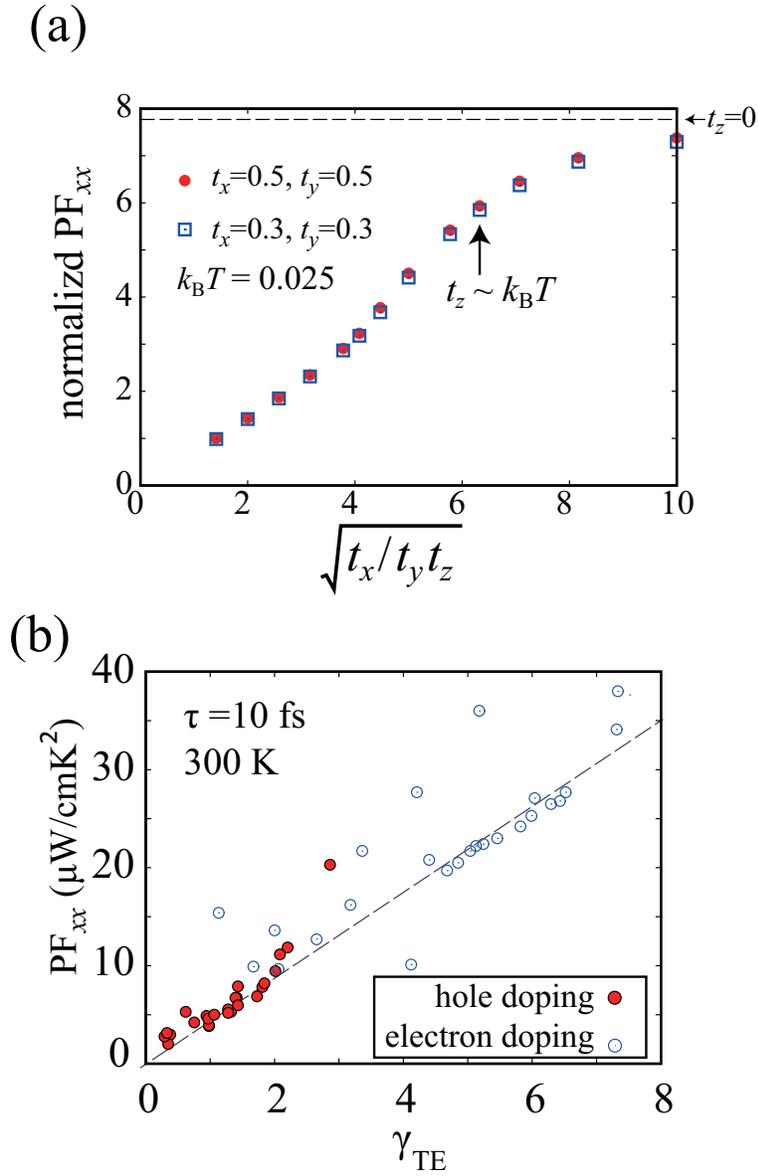}
  \caption{(a) The maximum value of the $xx$ component of the power factor plotted against $\sqrt{t_x/t_yt_z}$ and (b) the power factor plotted against $\gamma_{TE}$ at 300K.}
  \label{fig7}
  \end{center}
\end{figure}
 
The relationship between the band degeneracy $n_{TE}$ and the power factor shown in Fig. \ref{fig3_1} exhibits a positive correlation, but some results cannot be understood using only $n_{TE}$.
For example, the power factor at $n_{TE} \sim 1$ is widely spread from 3 to 15$\mu$W/cmK$^2$.
Moreover, the power factor for n-type ZnMg$_2$As$_2$ ($n_{TE} \sim 6$) is not larger than that of the compounds at $n_{TE} \sim 3$.
This is because we do not include the effect of the density of states and the group velocity, which mainly contribute to the thermoelectric effect within the constant relaxation time approximation.
The coexistence of large density of states and large group velocity is realized by strong anisotropy of the band structure.
Assuming the constant relaxation time approximation, the $xx$ component of $\bm{\sigma}(E)$ in Eq. (\ref{eq:sigma}) can be rewritten as
\begin{eqnarray}
  \sigma_{xx}(E) = \tau \bar{v}^{2}_{x}(E) D(E),
\end{eqnarray}
where $\bar{v}_{x}$ is the average of the group velocity at the energy $E$, and $D(E)$ is the density of states. 

When the band structure has a three dimensional character, namely, the band width along the $k_{x}, k_y$ and $k_z$ directions is much larger than $k_BT$, the density of states and the average value of the group velocity are proportional to $\sqrt{m_{x}m_{y}m_{z}}$ and $1/m_{x}$, respectively. $\sigma(E)$ is thus proportional to $\sqrt{m_{y}m_{z}/m_{x}}$, so that we introduce a quantity $\gamma_{TE}$ which simultaneously takes into account the anisotropy of the band structure and the valley degeneracy,
\begin{eqnarray}
  \gamma_{TE} = 4k_BT \sum_{i,\bm{k}} n_{\bm{k}}\sqrt{\frac{m_{i,\bm{k},y}m_{i,\bm{k},z}}{m_{i,\bm{k},x}}} \left( -\frac{df}{dE} \right)_{E=\Delta_{pi\bm{k}}/2}, \label{gamma_te}
\end{eqnarray}
where $m_{i,\bm{k},x/y/z}$ is the effective mass of the $i$-th band centered at the wave vector $\bm{k}$.\cite{comment2}

The above formula for $\gamma_{TE}$ is based on the assumption that the density of states is proportional to $\sqrt{m_x m_y m_z}$, 
which is not correct for electronic structures with low dimensionality. This can be understood by taking the $m_z \rightarrow \infty$ limit, where $\gamma_{TE} \rightarrow \infty$. In order to see what kind of correction should be made for  $\gamma_{TE}$ for low dimensional systems, we analyze a simple tightbinding model given as, 
\begin{eqnarray}
  \varepsilon(\bm{k}) = - 2t_x {\rm cos}(k_{x}a) - 2t_{y} {\rm cos}(k_{y}a) - 2t_{z} {\rm cos}(k_{z}c),
\end{eqnarray}
where $t_{x}, t_y $, and $t_z$ are the nearest neighbor hopping integrals along the $x$, $y$ and $z$ axis, respectively. 
Within this tightbinding model, the effective mass is expressed as $m_{x/y} \sim \hbar^2/(2t_{x/y}a^2)$ and $m_{z} \sim \hbar^2/(2t_{z}c^2)$.

The power factor obtained using this model is normalized such that the value at $t_x = t_y = t_y = 0.5$, and $k_BT$ = 0.025 is equal to 1.
The normalized power factor of this model is plotted against $\sqrt{t_x/t_yt_z} \propto \sqrt{m_ym_z/m_x}$ in Fig. \ref{fig7}(a).
Here we fix $t_x=t_y$ at 0.3 or 0.5 and vary $t_z$. 
The power factor for $t_{x}=t_{y}=0.5$ and $0.3$ exhibits essentially the same values for a fixed $t_z$ because $\sigma_{xx}(E)$ depends not on the value of $t_x$ and $t_y$ but on the ratio of $t_x$ to $t_y$.\cite{comment1}
The calculated power factor is almost proportional to $\sqrt{m_{y}m_{z}/m_{x}}$, but at around $t_z \lesssim k_{B}T$, the slope of the power factor as a function of $\sqrt{t_yt_z/t_x}$ becomes small.
This behavior of the power factor originates from the small band width along the $k_{z}$ axis, which is evaluated as $4t_{z}$.
Therefore, when $t_{z} \sim \hbar^2/(2m_{z}c^2) \lesssim  k_{B}T$, namely, $m_{z} \gtrsim 3m_{e}$ at 300K, the band structure should be regarded as a two-dimensional one as far as the power factor is concerned.
Based on the present analysis, in Eq. (\ref{gamma_te}), we set $m_z$ as $5m_e$ when $m_z > 5m_e$ because the power factor is essentially independent of $m_z$ for nearly two-dimensional systems with such large values of $m_z$. 

In Fig. \ref{fig7}(b), 
we plot the maximum value of the $xx$ component of the power factor for the 25 compounds against $\gamma_{TE}$, which clearly shows that the power factor is basically proportional to $\gamma_{TE}$.
Thus $\gamma_{TE}$ can be used to find promising thermoelectric materials.
On the other hand, we may say that the materials which largely deviate from the linear plot toward the larger $PF_{xx}$ regime in Fig. \ref{fig7} can be considered as those whose power factor is boosted owing presumably to the invalidity of the effective mass approximation.

%The power factor of some compounds deviates from the linear line, which means that these compounds cannot be understood within the effective mass approximation.
%By investigating the relationship between the theoretically obtained power factor and $\gamma_{TE}$, we could find thermoelectric materials which cannot be understood within the usual approximation.

\begin{figure}
  \begin{center}
  \includegraphics[width=8.5cm]{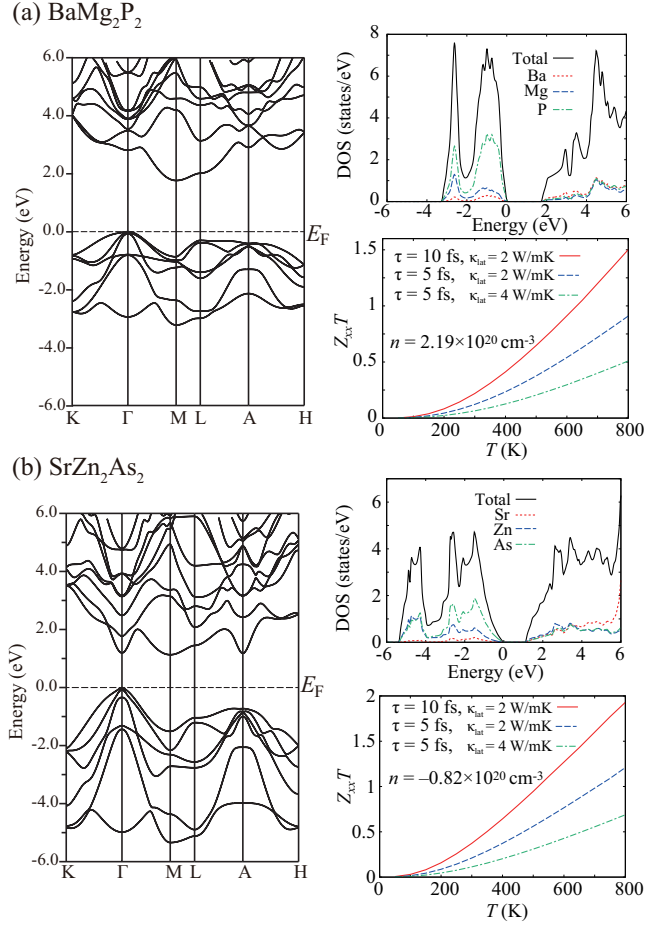}
  \caption{The electronic band structure, the density of states and the $xx$ component of the dimensionless figure of merit of (a) BaMg$_2$P$_2$ and (b) SrZn$_2$As$_2$.}
  \label{fig6}
  \end{center}
\end{figure}

Finally, in Fig. \ref{fig6}, we analyze in detail the two compounds which possess the largest largest dimensionless figure of merit in the electron and hole doped regime, respectively.
p-type BaMg$_2$P$_2$ exhibits $PF_{xx} \sim 20\mu$W/cmK$^2$ and $ZT > 0.2$ at 300K assuming the relaxation time to be 10 fs and the lattice thermal conductivity to be 2 W/mK.
It can be seen that $\gamma_{TE}$ of BaMg$_2$P$_2$ deviates from the linear line depicted in Fig. \ref{fig7}(b), which means that the valence band structure of BaMg$_2$P$_2$ cannot be described within the effective mass approximation.
The band structure shown in Fig. \ref{fig6}(a) actually indicates that the effective mass at the $\Gamma$ point is totally different from that at the $A$ point.
The band structure along A-L has nearly flat portion, which gives rise to the large density of states around the Fermi level.
We also show the temperature dependence of the dimensionless figure of merit at the doping level at which the dimensionless figure of merit is maximized at 300K.
The dimensionless figure of merit can be rewritten as $ZT = (\sigma/\tau)S^2T/(\kappa_{el}/\tau + \kappa_{lat}/\tau)$.
The terms of $\sigma/\tau$ and $\kappa_{el}/\tau$ do not depend on the relaxation time $\tau$ because it implicitly appears $K_m$ in the expression of $\bm{\sigma}$ and $\kappa_{el}$, so that $\kappa_{lat}/\tau$ is one of the key parameters that determine $ZT$.
In reality, the relaxation time $\tau$ tends to decrease when the temperature increases, but at the same time $\kappa_{lat}$ decreases,   
so that assuming a constant $\kappa_{lat}/\tau$ can be considered as a reasonable approximation.
The experimentally observed lattice thermal conductivity for the Sb and As systems is around 2 W/mK at 300K\cite{HZhang2008, Cao2010, Yu2008, Kihou2017, Kunioka2018, Kunioka2020}, so that $\kappa_{lat}$ with a similar order could be expected also in the P system, although the lattice thermal conductivity tends to be larger in the phosphides than in the antimonides.
For the n-type compounds, largest dimensionless figure of merit is obtained in SrZn$_2$As$_2$, which possesses larger PF of $\sim 35\mu$W/cmK$^2$ and large $ZT$ of 0.35 at 300K.
This is because the effective degree of the valley degeneracy is about 4 due to the conduction bands at the $\Gamma$ and M points.
Both the conduction band structures at the $\Gamma$ and M points exhibit two-dimensional shape, which results in the large density of states above the Fermi level.
The dimensionless figure of merit thus exceeds 1 at around 500 K.

\section{Conclusion}
We have theoretically investigated the thermoelectric properties of the 25 CaAl$_2$Si$_2$-type Zintl phase compounds $AB_2X_2$.
In order to analyze the relationship between the electronic band structure and the thermoelectric properties, we have introduced the quantity $n_{TE}$, which reflects the valley degeneracy.
We can see the positive correlation between $n_{TE}$ and the power factor.
It is found that the valley degeneracy is maximized when $X={\rm P}$ in the hole doped regime, and $A = {\rm Mg}$ or ${\rm Zn}$ in the electron doped regime.
We have also introduced $\gamma_{TE}$, which takes into account the effect of the effective mass and the valley degeneracy contributing to the thermoelectric properties.
We have found that the power factor of many of the 25 compounds is proportional to $\gamma_{TE}$.
Assuming the relaxation time to be 10 fs and the lattice thermal conductivity to be 2 W/mK, SrZn$_2$As$_2$ can exceed $ZT > 0.3$ at 300K for n-type, and $ZT$ of BaMg$_2$P$_2$ is larger than 0.2 at 300K for p-type.
Other compounds also exhibit large degree of the valley degeneracy, so that we can conclude that there is a room to find promising thermoelectric materials in the CaAl$_2$Si$_2$-type Zintl phase pnictides.

\begin{acknowledgment}

%\acknowledgment
We acknowledge C.H. Lee, K. Kihou, H. Kunioka and Y. Kimura for fruitful discussions.
This work was supported by JSPS KAKENHI (Grants No. 19K15436) and JST CREST (Grant No. JPMJCR20Q4), Japan.
\end{acknowledgment}

\begin{appendix}
\section{The results of the electronic band structure and the thermoelectric properties}
\begin{figure*}
  \begin{center} 
  \includegraphics[width=0.8\linewidth]{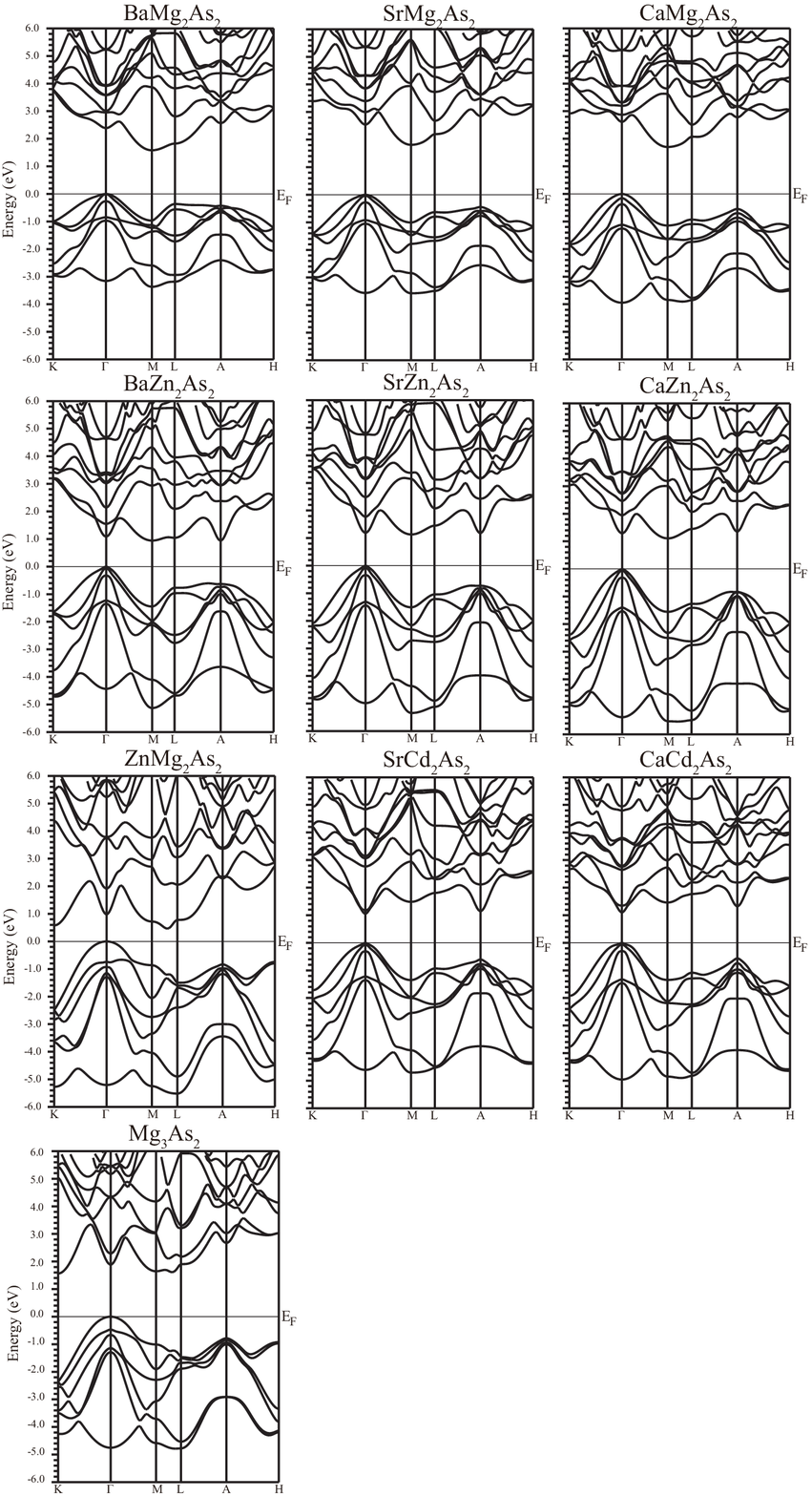}
  \caption{The electronic band structure of $AB_2$As$_2$}
  \label{fig_supp1}
  \end{center}
\end{figure*}

\begin{figure*}
  \begin{center} 
  \includegraphics[width=0.8\linewidth]{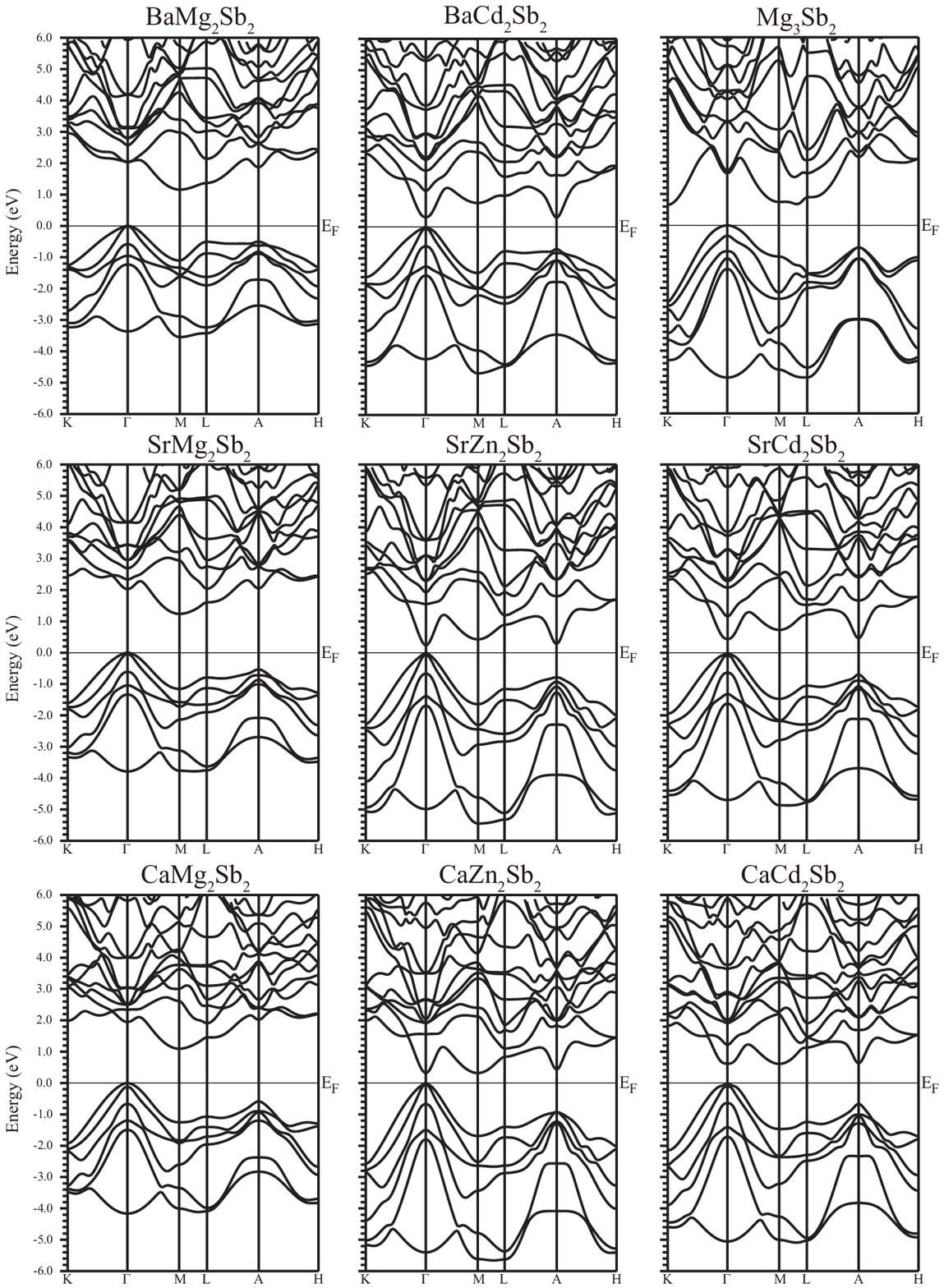}
  \caption{The electronic band structure of $AB_2$Sb$_2$}
  \label{fig_supp2}
  \end{center}
\end{figure*}

\begin{figure*}
  \begin{center} 
  \includegraphics[width=0.8\linewidth]{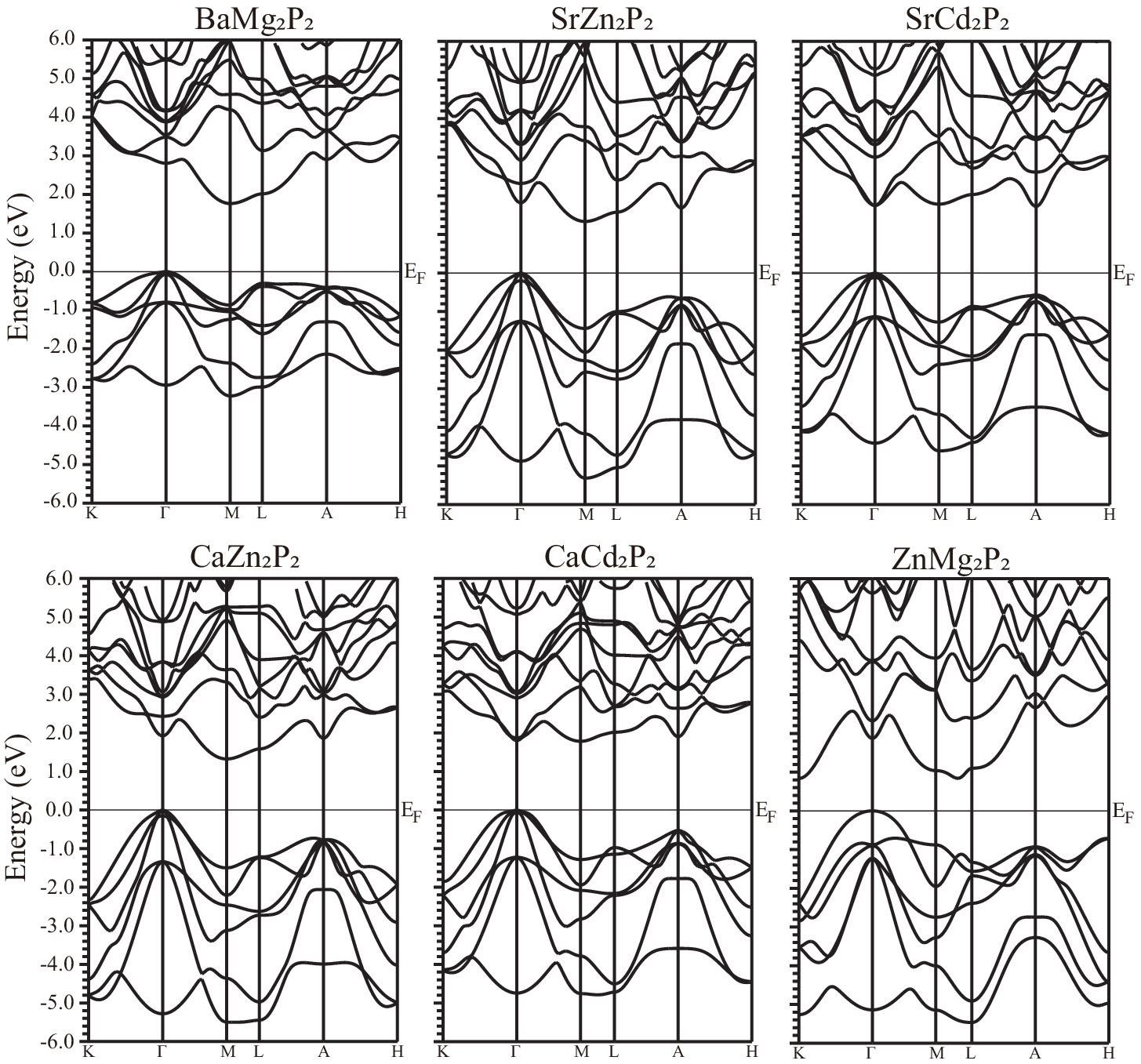}
  \caption{The electronic band structure of $AB_2$P$_2$}
  \label{fig_supp3}
  \end{center}
\end{figure*}

In this appendix, we show the electronic band structure and the thermoelectric properties of all the 25 compounds in Figs. \ref{fig_supp1}, \ref{fig_supp2} and \ref{fig_supp3}, and Tables \ref{table_sup1}, \ref{table_sup2} and \ref{table_sup3}.
The optimized lattice parameters $a$ and $c$, and the internal coordinates at the $B$ and $X$ sites are listed in Table \ref{table1} for the arsenides and Table \ref{table_sup1} for the 25 compounds.
The doping level at which the maximum power factor is obtained is listed in Tables \ref{table_sup2} and \ref{table_sup3} within the doping level of less than 0.1 holes/electrons at 300K.

\begin{table*}
  \begin{center}
    \caption{The lattice parameters $a$ and $c$ (unit: $\mbox{\AA}$), and the internal coordinates $z_B$ for the $B$ site and $z_X$ for the $X$ site, obtained by the first principles calculations. The internal coordinates are $1a (0,0,0)$ for the $A$ site, $2d (1/3, 2/3, z)$ for the atoms of the $B$ and $X$ sites. $n^{\rm hole}_{TE}$ and $n^{\rm ele}_{TE}$ are the effective degree of the valley degeneracy contributing to the thermoelectric effect at 300K in the electron and hole doped regime, respectively. \label{table_sup1}}
    \begin{tabular}[t]{|c|c|c|c|c|c|c|c|c|} \hline
                & $a$  & $c$  & $z_B$ & $z_X$ & $\Delta_{p1}$ & $\Delta_{p2}$ & $n^{\rm hole}_{TE}$ & $n^{\rm ele}_{TE}$ \\ \hline \hline
                   CaZn$_2$P$_2$ & 3.99 & 6.74 & 0.633 & 0.262 & 0.03 & 0.15 & 2.14  & 3.00  \\ \hline
                   CaCd$_2$P$_2$ & 4.25 & 6.94 & 0.637 & 0.238 & 0.02 & 0.07 & 2.62  & 4.50  \\ \hline
                   ZnMg$_2$P$_2$ & 4.11 & 6.26 & 0.652 & 0.222 & 0.87 & 0.91 & 1.00  & 7.44  \\ \hline
                   SrZn$_2$P$_2$ & 4.05 & 7.02 & 0.630 & 0.274 & 0.03 & 0.20 & 2.01  & 3.00  \\ \hline
                   SrCd$_2$P$_2$ & 4.32 & 7.20 & 0.635 & 0.252 & 0.03 & 0.12 & 2.24  & 3.97  \\ \hline
                   BaMg$_2$P$_2$ & 4.35 & 7.52 & 0.624 & 0.275 & 0.01 & 0.08 & 2.59  & 3.00  \\ \hline
                    Mg$_3$As$_2$ & 4.26 & 6.69 & 0.642 & 0.229 & 0.47 & 0.66 & 1.00  & 7.97  \\ \hline
                  CaMg$_2$As$_2$ & 4.34 & 7.04 & 0.635 & 0.247 & 0.15 & 0.37 & 1.20  & 3.00  \\ \hline
                  CaZn$_2$As$_2$ & 4.13 & 6.92 & 0.631 & 0.257 & 0.06 & 0.31 & 1.74  & 3.11  \\ \hline
                  CaCd$_2$As$_2$ & 4.38 & 7.09 & 0.633 & 0.234 & 0.06 & 0.30 & 1.71  & 1.04  \\ \hline
                  ZnMg$_2$As$_2$ & 4.23 & 6.49 & 0.646 & 0.218 & 0.74 & 0.93 & 1.00  & 6.17  \\ \hline
                  SrMg$_2$As$_2$ & 4.40 & 7.34 & 0.630 & 0.260 & 0.03 & 0.27 & 1.95  & 3.00  \\ \hline
                  SrZn$_2$As$_2$ & 4.19 & 7.19 & 0.629 & 0.270 & 0.08 & 0.34 & 1.57  & 3.84  \\ \hline
                  SrCd$_2$As$_2$ & 4.44 & 7.36 & 0.632 & 0.249 & 0.07 & 0.30 & 1.68  & 1.28  \\ \hline
                  BaMg$_2$As$_2$ & 4.47 & 7.68 & 0.625 & 0.272 & 0.00 & 0.26 & 2.02  & 3.00  \\ \hline
                  BaCd$_2$As$_2$ & 4.51 & 7.64 & 0.630 & 0.262 & 0.07 & 0.31 & 1.68  & 1.21  \\ \hline
                    Mg$_3$Sb$_2$ & 4.55 & 7.19 & 0.632 & 0.224 & 0.33 & 0.83 & 1.01  & 7.43  \\ \hline
                  CaMg$_2$Sb$_2$ & 4.64 & 7.48 & 0.630 & 0.241 & 0.11 & 0.66 & 1.35  & 3.00  \\ \hline
                  SrMg$_2$Sb$_2$ & 4.70 & 7.75 & 0.629 & 0.254 & 0.04 & 0.61 & 1.88  & 3.00  \\ \hline
                  BaMg$_2$Sb$_2$ & 4.77 & 8.09 & 0.626 & 0.267 & 0.00 & 0.58 & 2.00  & 3.00  \\ \hline
                  CaZn$_2$Sb$_2$ & 4.41 & 7.36 & 0.631 & 0.256 & 0.05 & 0.66 & 1.80  & 4.00  \\ \hline
                  SrZn$_2$Sb$_2$ & 4.48 & 7.63 & 0.632 & 0.270 & 0.07 & 0.66 & 1.64  & 1.27  \\ \hline
                  CaCd$_2$Sb$_2$ & 4.64 & 7.51 & 0.630 & 0.236 & 0.11 & 0.64 & 1.37  & 2.00  \\ \hline
                  SrCd$_2$Sb$_2$ & 4.70 & 7.75 & 0.631 & 0.249 & 0.09 & 0.63 & 1.49  & 1.04  \\ \hline
                  BaCd$_2$Sb$_2$ & 4.78 & 8.00 & 0.632 & 0.262 & 0.07 & 0.61 & 1.69  & 2.00  \\ \hline
    \end{tabular}
  \end{center}
\end{table*}

\begin{table*}
  \begin{center}
    \caption{The maximum value of the $xx$ and $zz$ components of the power factor $PF$ (unit: $\mu$W/cmK$^2$) and the dimensionless figure of merit $ZT$ obtained with the relaxation time to be 10 fs and $\kappa_{lat} = 2$ W/mK within the doping level of less than 0.1 holes at 300K. $p$ is the amount of the hole doping per unit cell at which the power factor or the dimensionless figure of merit is maximized at 300K. \label{table_sup2}}
\begin{tabular}[t]{|c|cc|cc|cc|cc|} \hline
                & $p$ & $PF_{xx}$ & $p$ & $PF_{zz}$ & $p$ & $Z_{xx}T$ & $p$ & $Z_{zz}T$ \\ \hline \hline
  CaZn$_2$P$_2$ & 0.037 &  7.8 & 0.027 &  8.5 & 0.004 & 0.10 & 0.020 & 0.09 \\ \hline
  CaCd$_2$P$_2$ & 0.023 & 11.8 & 0.013 & 13.7 & 0.014 & 0.15 & 0.009 & 0.18 \\ \hline
  ZnMg$_2$P$_2$ & 0.097 &  3.0 & 0.004 & 13.7 & 0.012 & 0.03 & 0.003 & 0.18 \\ \hline
  SrZn$_2$P$_2$ & 0.007 &  8.2 & 0.050 &  7.9 & 0.004 & 0.11 & 0.035 & 0.07 \\ \hline
  SrCd$_2$P$_2$ & 0.044 & 11.1 & 0.029 & 11.2 & 0.022 & 0.11 & 0.020 & 0.13 \\ \hline
  BaMg$_2$P$_2$ & 0.070 & 20.3 & 0.077 & 17.9 & 0.027 & 0.22 & 0.027 & 0.20 \\ \hline
   Mg$_3$As$_2$ & 0.098 &  2.0 & 0.003 & 10.6 & 0.009 & 0.03 & 0.009 & 0.14 \\ \hline
 CaMg$_2$As$_2$ & 0.097 &  5.3 & 0.002 &  5.8 & 0.027 & 0.05 & 0.002 & 0.08 \\ \hline
 CaZn$_2$As$_2$ & 0.082 &  5.3 & 0.071 &  5.4 & 0.005 & 0.07 & 0.005 & 0.06 \\ \hline
 CaCd$_2$As$_2$ & 0.099 &  6.7 & 0.009 &  8.6 & 0.012 & 0.07 & 0.007 & 0.12 \\ \hline
 ZnMg$_2$As$_2$ & 0.099 &  2.8 & 0.003 & 13.0 & 0.014 & 0.03 & 0.002 & 0.17 \\ \hline
 SrMg$_2$As$_2$ & 0.097 &  6.9 & 0.014 &  6.6 & 0.010 & 0.09 & 0.010 & 0.09 \\ \hline
 SrZn$_2$As$_2$ & 0.097 &  5.5 & 0.094 &  5.4 & 0.007 & 0.07 & 0.007 & 0.05 \\ \hline
 SrCd$_2$As$_2$ & 0.097 &  6.7 & 0.097 &  6.3 & 0.008 & 0.07 & 0.008 & 0.07 \\ \hline
 BaMg$_2$As$_2$ & 0.095 &  9.5 & 0.095 &  8.7 & 0.013 & 0.11 & 0.013 & 0.10 \\ \hline
 BaCd$_2$As$_2$ & 0.099 &  7.9 & 0.099 &  7.1 & 0.010 & 0.08 & 0.011 & 0.07 \\ \hline
   Mg$_3$Sb$_2$ & 0.083 &  3.1 & 0.002 &  8.1 & 0.071 & 0.03 & 0.002 & 0.11 \\ \hline
 CaMg$_2$Sb$_2$ & 0.016 &  4.2 & 0.002 &  4.1 & 0.014 & 0.05 & 0.002 & 0.06 \\ \hline
 SrMg$_2$Sb$_2$ & 0.009 &  5.2 & 0.010 &  5.2 & 0.007 & 0.07 & 0.007 & 0.07 \\ \hline
 BaMg$_2$Sb$_2$ & 0.009 &  5.9 & 0.010 &  5.7 & 0.007 & 0.08 & 0.007 & 0.08 \\ \hline
 CaZn$_2$Sb$_2$ & 0.004 &  3.8 & 0.004 &  4.0 & 0.003 & 0.05 & 0.003 & 0.06 \\ \hline
 SrZn$_2$Sb$_2$ & 0.006 &  3.9 & 0.005 &  3.8 & 0.004 & 0.05 & 0.004 & 0.05 \\ \hline
 CaCd$_2$Sb$_2$ & 0.012 &  4.9 & 0.006 &  7.2 & 0.011 & 0.06 & 0.006 & 0.10 \\ \hline
 SrCd$_2$Sb$_2$ & 0.010 &  4.6 & 0.007 &  5.4 & 0.009 & 0.06 & 0.005 & 0.07 \\ \hline
 BaCd$_2$Sb$_2$ & 0.008 &  5.0 & 0.006 &  5.0 & 0.006 & 0.07 & 0.005 & 0.07 \\ \hline
 \end{tabular}
\end{center}
\end{table*}

\begin{table*}
  \begin{center}
    \caption{The maximum value of the $xx$ and $zz$ components of the power factor $PF$ (unit: $\mu$W/cmK$^2$) and the dimensionless figure of merit $ZT$ obtained with the relaxation time to be 10 fs and $\kappa_{lat} = 2$ W/mK within the doping level of less than 0.1 electrons at 300K. $n$ is the amount of the electron doping per unit cell at which the power factor or the dimensionless figure of merit is maximized at 300K. \label{table_sup3}}
\begin{tabular}[t]{|c|cc|cc|cc|cc|} \hline  
                 & $n$ & $PF_{xx}$ & $n$ & $PF_{zz}$ & $n$ & $Z_{xx}T$ & $n$ & $Z_{zz}T$ \\ \hline \hline
   CaZn$_2$P$_2$ & 0.016 & 26.5 & 0.008 & 3.3  & 0.007 & 0.30 & 0.008 & 0.05 \\ \hline
   CaCd$_2$P$_2$ & 0.054 & 38.0 & 0.028 & 5.2  & 0.012 & 0.35 & 0.016 & 0.07 \\ \hline
   ZnMg$_2$P$_2$ & 0.091 & 20.8 & 0.012 & 16.9 & 0.007 & 0.23 & 0.006 & 0.20 \\ \hline
   SrZn$_2$P$_2$ & 0.017 & 25.3 & 0.008 & 3.3  & 0.007 & 0.29 & 0.008 & 0.05 \\ \hline
   SrCd$_2$P$_2$ & 0.034 & 27.7 & 0.025 & 5.9  & 0.010 & 0.27 & 0.022 & 0.08 \\ \hline
   BaMg$_2$P$_2$ & 0.020 & 19.7 & 0.013 & 3.9  & 0.010 & 0.24 & 0.011 & 0.06 \\ \hline
    Mg$_3$As$_2$ & 0.037 & 36.0 & 0.012 & 19.8 & 0.014 & 0.33 & 0.007 & 0.25 \\ \hline
  CaMg$_2$As$_2$ & 0.019 & 27.1 & 0.062 & 5.7  & 0.011 & 0.32 & 0.035 & 0.07 \\ \hline
  CaZn$_2$As$_2$ & 0.083 & 24.2 & 0.008 & 3.7  & 0.006 & 0.27 & 0.008 & 0.05 \\ \hline
  CaCd$_2$As$_2$ & 0.097 & 15.4 & 0.049 & 5.9  & 0.063 & 0.07 & 0.039 & 0.06 \\ \hline
  ZnMg$_2$As$_2$ & 0.099 & 16.2 & 0.006 & 13.8 & 0.005 & 0.18 & 0.004 & 0.18 \\ \hline
  SrMg$_2$As$_2$ & 0.015 & 22.4 & 0.031 & 4.6  & 0.008 & 0.27 & 0.018 & 0.06 \\ \hline
  SrZn$_2$As$_2$ & 0.015 & 34.1 & 0.010 & 3.7  & 0.009 & 0.37 & 0.010 & 0.05 \\ \hline
  SrCd$_2$As$_2$ & 0.097 & 9.9  & 0.097 & 3.3  & 0.003 & 0.13 & 0.013 & 0.04 \\ \hline
  BaMg$_2$As$_2$ & 0.019 & 20.5 & 0.012 & 3.7  & 0.011 & 0.25 & 0.011 & 0.05 \\ \hline
  BaCd$_2$As$_2$ & 0.098 & 13.6 & 0.084 & 3.4  & 0.002 & 0.11 & 0.076 & 0.04 \\ \hline
    Mg$_3$Sb$_2$ & 0.056 & 27.7 & 0.011 & 14.2 & 0.008 & 0.22 & 0.006 & 0.18 \\ \hline
  CaMg$_2$Sb$_2$ & 0.021 & 22.2 & 0.013 & 3.5  & 0.009 & 0.26 & 0.011 & 0.05 \\ \hline
  SrMg$_2$Sb$_2$ & 0.018 & 21.7 & 0.018 & 3.8  & 0.009 & 0.26 & 0.014 & 0.05 \\ \hline
  BaMg$_2$Sb$_2$ & 0.024 & 23.0 & 0.014 & 3.1  & 0.012 & 0.27 & 0.012 & 0.05 \\ \hline
  CaZn$_2$Sb$_2$ & 0.014 & 26.8 & 0.012 & 4.4  & 0.006 & 0.31 & 0.009 & 0.06 \\ \hline
  SrZn$_2$Sb$_2$ & 0.049 & 12.7 & 0.030 & 3.9  & 0.002 & 0.14 & 0.028 & 0.05 \\ \hline
  CaCd$_2$Sb$_2$ & 0.017 & 21.7 & 0.031 & 4.9  & 0.009 & 0.23 & 0.026 & 0.07 \\ \hline
  SrCdSb$_2$$_2$ & 0.003 & 9.6  & 0.072 & 3.7  & 0.002 & 0.13 & 0.066 & 0.05 \\ \hline
  BaCd$_2$Sb$_2$ & 0.002 & 10.1 & 0.099 & 3.0  & 0.002 & 0.14 & 0.095 & 0.04 \\ \hline
\end{tabular}
\end{center}
\end{table*}

\end{appendix}


\begin{thebibliography}{9}
\bibitem{comment2}
At $\bm{k}={\rm M}$, we replace the term of the effective mass from $\sqrt{m_ym_z/m_x}$ to ($\sqrt{m_ym_z/m_x}+\sqrt{m_xm_z/m_y})/2$ because of the symmetry of the band structure centered at the M point.
Generally, the band structure having three- or six-fold symmetry cannot be reproduced within the effective mass approximation using $m_x$ and $m_y$, but we used this approximation in this study because not $m_x$ and $m_y$ but $m_z$ and $n_{k}$ are important factors for $\gamma_{TE}$ in the 1-2-2 type Zintl phase compunds.
\bibitem{comment1}
Even in the two-dimensional system, namely, in the case of the $t_z \rightarrow 0$ limit, the density of states and the average value of the group velocity are proportional to $\sqrt{m_{x}m_{y}}$ and $1/m_{x}$, respectively, so that $\sigma(E) \propto \sqrt{m_{y}/m_x}$ is determined by the ratio of $m_x$ to $m_y$.
\end{thebibliography}
\end{document}